\documentclass[aps,prl,twocolumn,preprintnumbers,amsmath,amssymb,10pt]{revtex4}
\usepackage{amsmath,graphicx}

\begin{document}
\title{Exploring Inelasticity in the S-Matrix Bootstrap
}
\author{Ant\'onio Antunes$^{a,b}$, Miguel S. Costa$^a$, Jos\'e Pereira$^a$} 
\preprint{DESY-23-015}
\affiliation{
$^a$  Centro de F\'isica do Porto, Departamento de F\'isica e Astronomia,
Faculdade de Ci\^encias da Universidade do Porto,
Rua do Campo Alegre 687, 4169-007 Porto, Portugal\\
$^b$ Deutsches Elektronen-Synchrotron DESY, Notkestr. 85, 22607 Hamburg, Germany
}

\begin{abstract}
The modern S-Matrix Bootstrap provides non-perturbative bounds on low-energy aspects of scattering amplitudes, leveraging the constraints of unitarity, analyticity and crossing. Typically, the solutions saturating such bounds also saturate the unitarity constraint as much as possible, meaning that they are almost exclusively elastic. This is expected to be unphysical in $d>2$ because of Aks' theorem. We explore this issue by adding inelasticity as an additional input, both using a primal approach in general dimensions which extends the usual ansatz, and establishing a dual formulation in the 2d case. We then measure the effects on the low-energy observables where we observe stronger bounds than in the standard setup.
\end{abstract}
\maketitle
\nopagebreak

{\bf Introduction.}
Scattering amplitudes are some of the most studied observables in quantum field theory. They encode the probability amplitudes of transitions between asymptotic states with a definite number of particles. The simplest non-trivial amplitude $_{\textrm{in}}\langle p_1,p_2|p_3,p_4\rangle_{\textrm{out}}$, corresponding to 2-2 scattering has been extensively studied for decades, notably through Feynman perturbation theory, which extracts the connected amplitude through the LSZ procedure which takes as input a four-point correlation function. Additionally, modern on-shell perturbative techniques formulate these amplitudes in terms of simpler objects and give illuminating insights on the structure of the answers \cite{Travaglini:2022uwo}. However, these methods are of limited applicability and general methods to tackle the non-perturbative case are lacking. 

A glimmer of hope is provided by the S-matrix Bootstrap, which takes the 2-2 amplitude and imposes severe constraints on its form by imposing elementary properties that should hold non-perturbatively: unitarity, analyticity and crossing symmetry \cite{Paulos:2016but,Paulos:2017fhb,Homrich:2019cbt,Doroud:2018szp,He:2018uxa,Cordova:2018uop,Paulos:2018fym,Guerrieri:2018uew,Cordova:2019lot,Bercini:2019vme,Gabai:2019ryw,EliasMiro:2019kyf,Karateev:2019ymz,Guerrieri:2020kcs,Guerrieri:2020bto,Bose:2020shm,Correia:2020xtr,Hebbar:2020ukp,He:2021eqn,Chen:2021pgx,EliasMiro:2021nul,Guerrieri:2021tak,Guerrieri:2021ivu,Tourkine:2021fqh,Correia:2022dcu,Correia:2022dyp,EliasMiro:2022xaa,Haring:2022sdp,Kruczenski:2022lot,Karateev:2022jdb,Chen:2022nym,Guerrieri:2022sod}. To understand the space of possible scattering amplitudes, one shoots towards its boundaries, by maximizing along some finite-dimensional observable space. Natural candidates are the residues or cubic couplings $g$ associated to a bound-state of mass $m_b$, such that $T(s,t)\sim g^2/(s-m_b^2)$, or the low-energy expansion around the crossing symmetric point $s=t=u=4m^2/3$, where one defines the coefficients 
\begin{equation}
	\Lambda_{ab}=\partial_s^a\partial_t^b T(4/3,4/3,4/3)\,,
\end{equation} where we set $m=1$ and considered neutral scalar particles for simplicity, as we will throughout our analysis. Maximizing or minimizing such observables gives a finite-dimensional slice of the infinite-dimensional space of scattering amplitudes, but also yields explicit amplitudes saturating the bounds. Typically, such solutions saturate the unitarity constraints \cite{Paulos:2016but}. For example, in 2d, where we can write the full S-matrix, including the disconnected contribution as $S(s)=1+i T(s)/(\sqrt{s(s-4)})$, one finds solutions that satisfy
\begin{equation}
	|S(s)|\approx 1, \quad s\geq 4\,,
\end{equation}
saturating the unitarity condition $|S(s)|\leq1$ up to very high energies. This saturation is expected in the so-called \textit{elastic unitarity} region, where the only possible end-states are precisely the two-particle states one starts with. On the other hand, at large enough energies, new normal thresholds will emerge, for example at $s=(nm)^2$, corresponding to the $n-$particle threshold or at $4m_b^2$ corresponding to the continuum of two-bound state pairs. The fact that $|S(s)|=1$ implies that the amplitude is purely elastic, meaning that there is no particle production $S_{2\to X\neq2}=0$. This is, of course, the key property of integrable field theories in 2 dimensions, which evidently saturate the corresponding bounds. A first study of the effects of particle production in 2 dimensional S-Matrices (which does not use the traditional S-matrix Bootstrap, but instead a fixed point method) was done in \cite{Tourkine:2021fqh}.

In contrast, in higher dimensions, Aks' theorem ensures that non-trivial scattering implies non-vanishing particle production, albeit only asymptotically at large energies \cite{Aks:1965qga}. A quantitative lower bound on this production was derived by Roy and Martin \cite{Martin:2017ndt}, stating that
\begin{equation}
	\label{eq:RoyMartin}
	\sigma_{\textrm{inel}}(s)\gtrsim\frac{\text{const}}{s^{5/2}}\,\text{Exp}\left[-\frac{9\sqrt{s}}{8}\ln s\right].
\end{equation}

\begin{figure}
	\includegraphics[width=0.8\linewidth]{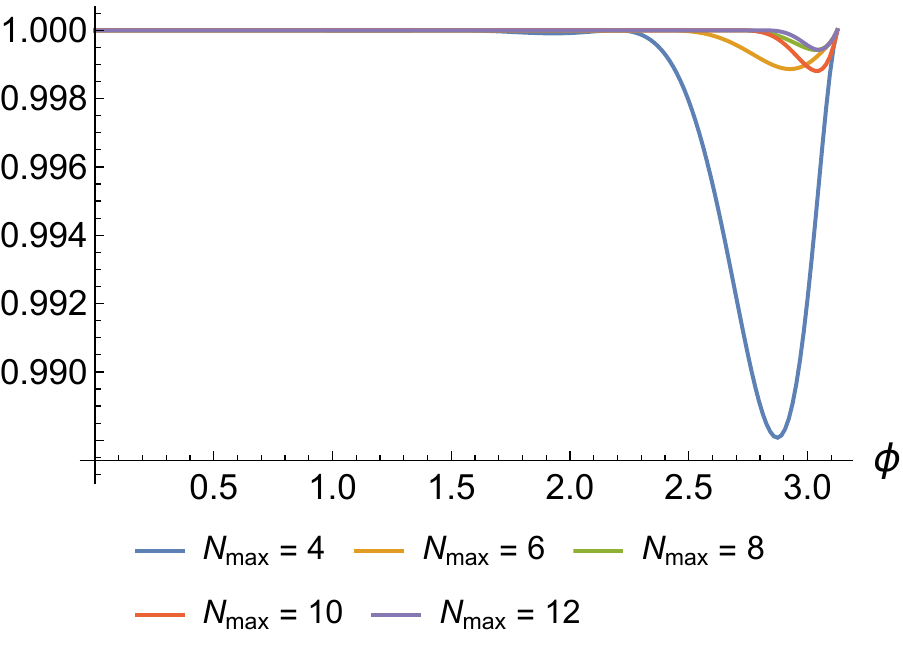}
	\caption[justification=justified]{The absolute value of $S_0$ for several $N_{max}$. $\phi$ is the argument of the complex $\rho$ variable defined below, such that $s$ correspondingly increases from $4$ to infinity.}
	\label{fig:S12comps}
\end{figure}

On the other hand, solutions to the higher dimensional S-Matrix bounds in practice also remain elastic up to very high energies, which is in tension (but not in sharp contradiction, since \cite{Aks:1965qga,Martin:2017ndt} make only asymptotic predictions) with the minimal particle production \cite{Paulos:2017fhb}. Decomposing onto partial amplitudes of well-defined angular momentum $\ell$, one observes
\begin{equation}
	|S_\ell(s)|\approx 1, \quad s\geq 4,
\end{equation}
as can be seen for example in figure \ref{fig:S12comps}, for the case of maximizing the "quartic coupling" $\Lambda_{00}$. 
Presumably, for the physics of inelasticity and particle production to be more easily captured by the bootstrap, the constraints associated to higher-point amplitudes should be included. For example, by including 3-particle states, one can require that the matrix of inner products between all 2 and 3 particle states is positive semi-definite. This would then impose non-trivial constraints between 4-, 5- and 6-point amplitudes. In practice however, technical control over higher-point amplitudes is very limited non-perturbatively and their analytic and crossing properties are an important open problem.

In this paper, we take a more pragmatic and exploratory approach. We consider the following question: Given some postulated inelasticity profile, how much do the S-matrix bounds actually change? We will see that the answer is certainly non-zero and we will quantify it in several examples. Along the way, we will develop an ansatz that correctly captures the additional thresholds and speculate on less standard maximization problems which might inherently capture particle production. 

There are a few motivations to do this. First, in gravitational scattering there is a natural behaviour for the inelastic part of the amplitude, stemming from black-hole production and subsequent Hawking radiation. Indeed, one expects the elastic part of the amplitude to be exponentially small at large $s$, as a consequence of the Bekenstein-Hawking formula \cite{Guerrieri:2022sod}. Secondly, in realistic particle-physics setups, as in pion scattering in QCD, one finds that the physical theories are often not so close to the bounds derived by the S-matrix Bootstrap \cite{Guerrieri:2018uew,Guerrieri:2020bto}. Information regarding the total cross-section, which of course captures the inelastic physics as well, could help bring QCD closer to the bound. Finally, in the simplest non-integrable QFT, $2d$ Ising field theory, recent advances in quantum spin chain simulations are on the brink of allowing us to study $2\to3$ scattering \cite{note4}. How such knowledge could strengthen the Bootstrap bounds and zoom in on a solution of the theory is an interesting question as well. 

{\bf Primal in 2d.}
We begin from the standard Mandelstam-like $\rho-$ansatz for a 2-dimensional S-matrix of $2-2$ scattering of mass $m$ particles with a bound state of mass $m_b$ \cite{Paulos:2017fhb}
\begin{equation}\label{eqn:elasticAnsatz}
	S(s) =\left[ \frac{-J_\rho\,g^2_{\rho}}{\rho(s)-\rho(m_b^2)}+ s\leftrightarrow t \right]+\sum_{a,b=0}^{N_{\textrm{max}}}c_{(ab)}\,\,\rho_{s}^a\rho_{t}^b
\end{equation}
where
\begin{equation}
\rho_s = \frac{\sqrt{4-s_0}-\sqrt{4-s}}{\sqrt{4-s_0}+\sqrt{4-s}},\quad s = \frac{s_0(1-\rho_s)^2+16\rho_s}{(1+\rho_s)^2}\,,
\end{equation}
This ansatz uses an analytic extension which assumes only the normal thresholds in $s>4m^2$ and $t>4m^2$ simultaneously. Each $\rho$ variable then maps the cut plane to a disk with the center at $s_0$ which is henceforth set to 2. Imposing the relation $s+t=4m^2$ yields a crossing-symmetric S-matrix with the expected analytic properties in the single variable $s$, but we will often write $t\equiv4m^2-s$ for convenience. To make it a full-fledged S-matrix, one imposes unitarity for physical scattering energies
\begin{equation}
	|S(s)|\leq1\,,  \quad\quad s\geq 4m^2\,.
\end{equation}
This statement follows from the elementary quantum mechanical principle of probability conservation
\begin{equation}
	\sum_X |S_{2\to X}|^2=1\,,
\end{equation}
which of course bounds the absolute value of the $2-2$ component $S(s)$ by 1. As mentioned above, integrable theories satisfy the above principle with a single term, the elastic component. Generic theories, e.g. $\phi^4$ theory, will admit for example a $S_{2\to 4}$ component which will then ensure that the absolute value of $S(s)$ is strictly less than 1. We parametrize this by an inelasticity profile $\beta(s)$, thereby refining the bootstrap problem to 
\begin{equation}
	|S(s)|^2\leq \beta(s),\quad s\geq 4m^2; \quad\beta(s)\geq 0;
\end{equation}
and we require the profile to have the following structure (we henceforth set $m^2=1$):
\begin{equation}\label{eqn:unitarityinelastic}
 \beta(s) =
	\begin{cases}
		1, &\quad 4\leq s\leq16\\
		\text{profile}(s) &\quad s>16\\
	\end{cases}
\end{equation}
We are hence trying to impose elastic unitarity, meaning we only allow particle production after the 4-particle threshold energy has been achieved. Of course, this choice is arbitrary and we can replace $16$ by $s^*$, a tunable threshold value, which could be a 3-particle or 2-bound state threshold. We will study the effect of this parameter later on. We will see that complying with elastic unitarity $|S(s)|=1$ for $s<s^*$ is challenging for the usual ansatz (\ref{eqn:elasticAnsatz}), and this will be one of the motivations to construct a more general parametrization.

We can now try to study standard maximization problems, for example, maximizing the cubic coupling to the bound state. The solution to this problem is given by the sine-Gordon breather S-matrix, but by imposing a non-trivial profile, we will explore the interior of this bound. For an alternative way to explore this interior see Appendix A. In 2-dimensions, we are blessed with an explicit solution to the S-matrix which reconstructs the phase given its absolute value \cite{Paulos:2016but,Tourkine:2021fqh}. It is given by
 \begin{equation}\label{eqn:analyticinelastic}
 \small
 	\frac{S}{S_{\text{el}}} = \exp \left[ \int_{4}^{\infty}\frac{ds'}{2\pi i}\sqrt{\frac{st}{s't'}}
 		\left(\frac{\log(\beta(s'))}{s-s'}+s \leftrightarrow t\right)\right] \,
 \end{equation}
and will allow us to check our results. We will of course not have such a luxury in four dimensions.
Let us consider then two simple profiles:
\begin{equation}\label{eqn:betaprofile1}
\small
	\beta^{(\textrm{s})} = \begin{cases}
		1, \,\, 4\leq s\leq s^*\\
		1-\alpha, \,\, s>s^*\\
			\end{cases}
		  \beta^{(\textrm{e})}  = \begin{cases}
			1, \,\, 4\leq s\leq s^*\\
	      e^{-\alpha\sqrt{s-s^*}} \,\, s>s^*\\
	\end{cases}
\end{equation}
While these are to some extent arbitrary functions, we believe these toy models should be qualitatively useful to understand the effects of more realistic inelasticity profiles. The first, a step function, is universally applicable as one can approximate any profile by a sum of such terms. In fact, it should even work quantitatively well near threshold, since it captures the particle production predicted from form-factor perturbation theory \cite{Zamolodchikov:2011wd}. The second profile, a sub-exponential decay (chosen to ensure convergence of \ref{eqn:analyticinelastic}) emulates the black-hole production patterns, relevant for example in the gravitational scattering processes studied in \cite{Guerrieri:2021ivu,Guerrieri:2022sod}.

\begin{figure}
		\includegraphics[width=0.8\linewidth]{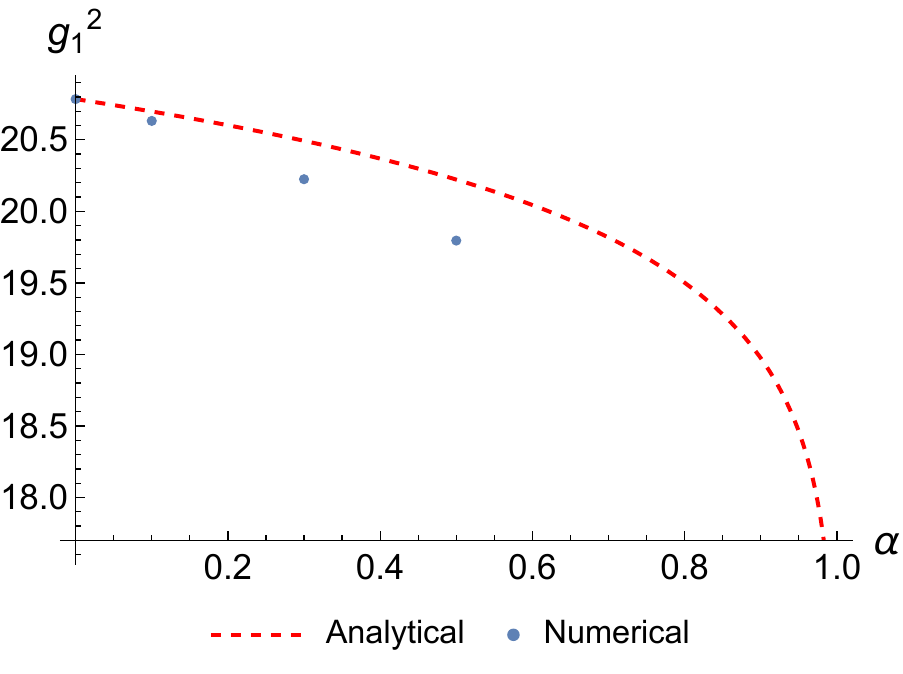}
		\caption{Direct comparison between profile $\beta^{(\textrm{s})}$ and the analytical results}
		\label{fig:numvsan2dinprimal}
	\end{figure}

Using the standard maximization algorithms with the additional $\beta$ profiles, we find the results of figure \ref{fig:numvsan2dinprimal}, which used $N_{\textrm{max}} = 5$ and a grid of 100 points for unitarity. 

\begin{figure}
	\includegraphics[width=0.8\linewidth]{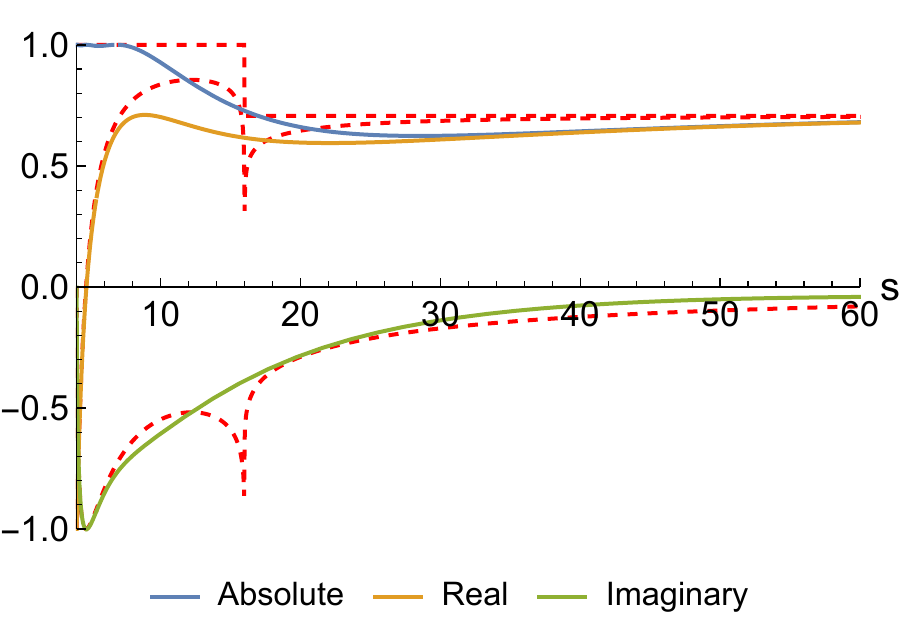}
	\caption{S-matrix components for the inelastic profile $\beta^{(\textrm{s})}$ with $\alpha=0.5$ and the analytical results in red dashes lines.}
	\label{fig:Smatcompsinel2d}
\end{figure}

Very clearly, the standard algorithm is under-performing. Looking with more detail into the numerical solution, we find the S-matrix components of figure \ref{fig:Smatcompsinel2d}.
It is easy to see that not only the threshold discontinuity is not correctly captured, but that it is also quite difficult to satisfy elastic unitarity and to saturate the inelasticity profile we imposed. Similar results are found for the other profile, and for other problems, for example extremization of $S(s=2)$ subject to the $\beta(s)$ constraints.

\begin{figure}
	\centering
	\includegraphics[width=.75\linewidth]{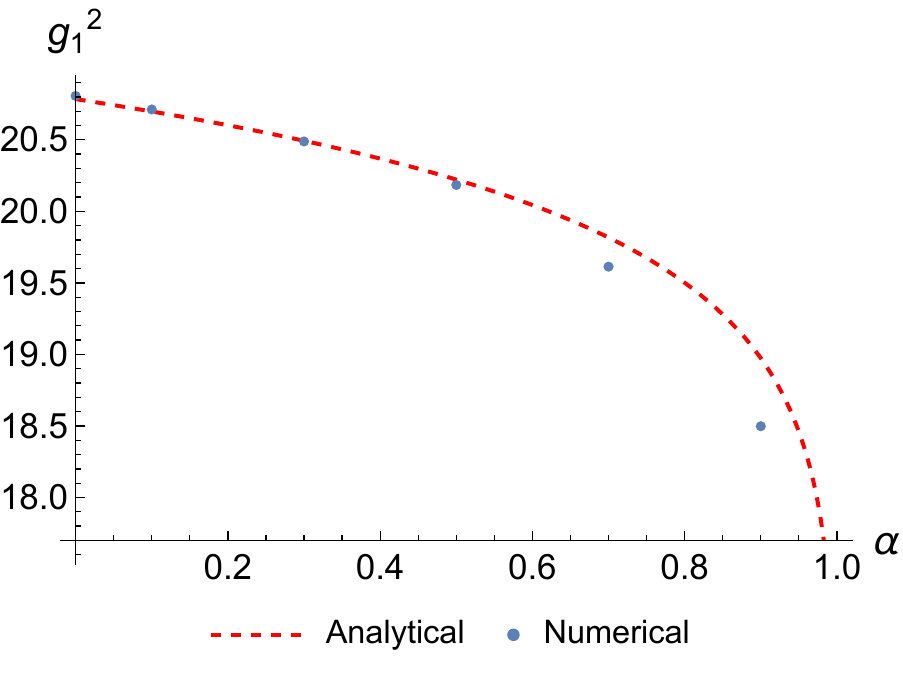}
	\caption[Analytical solution vs numerical data, 2D inelastic with $\rho_{16}$]{Direct comparison between the profile $\beta^{(\textrm{s})}$ and the analytical results. The bound-state mass is fixed, $m_1^2 = 3$, $N_{\textrm{max}} = 5$ and $\bar{N}_{\textrm{max}} = 2$.}
	\label{fig:numvsan2dinprimalb1rho16}
\end{figure}

\begin{figure}
	\centering
	\includegraphics[width=.8\linewidth]{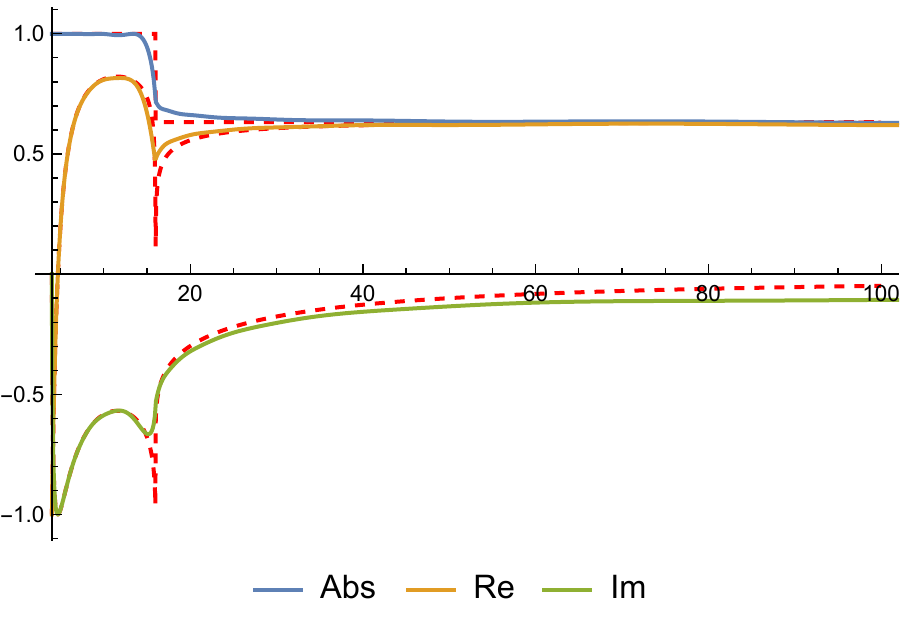}
	\caption[S-matrix components, 2D primal with $\rho_{16}$]{ S-matrix components for the inelastic profile $\beta^{(\textrm{s})}$ using the new Ansatz of (\ref{eqn:newinelasticAnsatz}). In red are the components of the analytical solution. Here, the parameters used were $m_1^2=3$, $N_{\textrm{max}} = 10$ and $\bar{N}_{\textrm{max}} = 10$ and $\alpha = 0.6$.}
	\label{fig:SmatcompsinelasticNewAnsatz}
\end{figure}

To capture the effects of the additional threshold, we first introduce a simple generalization of the $\rho$ variables, which map the cut plane starting at the new threshold $s^*$ to the disk. They manifestly satisfy the original analyticity assumptions and are given by
\begin{equation}\label{eqn:s16map}
	\bar{\rho}_{s} \equiv \frac{\sqrt{s^*-s_0}-\sqrt{s^*-s}}{\sqrt{s^*-s_0}+\sqrt{s^*-s}},
\end{equation}
It is then natural to extend the ansatz for the S-matrix to have the form
\begin{align}
	\label{eqn:newinelasticAnsatz}
		S_{\textrm{improved}}(s) = S(s) + \sum_{a,b=0}^{\bar{N}_{\textrm{max}}}d_{(ab)}\,\,\bar{\rho}_{s}^a\bar{\rho}_{t}^b\,,
\end{align}
where we introduced an independent cutoff $\bar{N}_{\textrm{max}}$.
As for the standard $\rho$ variable, the $\bar{\rho}$ satisfy the simple algebraic identity
\begin{equation}
	\bar{\rho}_s+\bar{\rho}_t+4\bar{\rho}_s\bar{\rho}_t+\bar{\rho}_s^2\bar{\rho}_t+\bar{\rho}_s\bar{\rho}_t^2=0\,,
\end{equation}
which allows us to eliminate several of the coefficients in (\ref{eqn:newinelasticAnsatz}). In fact we will keep exactly the same $a,b$ as we do for the coefficients in (\ref{eqn:elasticAnsatz}). It is easy to check that additional terms involving products of the different $\rho$'s and similar variables parametrizing a discontinuity between $4$ and $s^*$ are redundant, and don't further modify the bounds \cite{note1}. We can now repeat the cubic coupling maximization, yielding figure \ref{fig:numvsan2dinprimalb1rho16}.
Not only do the milder discontinuities get a significant enhancement, but also the sharper ones as $\alpha\to 1$ appear to be getting qualitatively reproduced. A look at the S-matrix extremizing the bounds, figure \ref{fig:SmatcompsinelasticNewAnsatz}, also reveals that the additional discontinuity is correctly captured, and elastic unitarity is satisfied until much closer to the threshold.
In fact, we improve the convergence by at least an order of magnitude. It turns out that this ansatz will work remarkably well in four dimensions, but we first establish even more accurate results in two dimensions by resorting to the dual formalism \cite{Guerrieri:2020kcs,He:2021eqn,Guerrieri:2021tak}.

\begin{figure}
	\centering
		\includegraphics[width=0.7\linewidth]{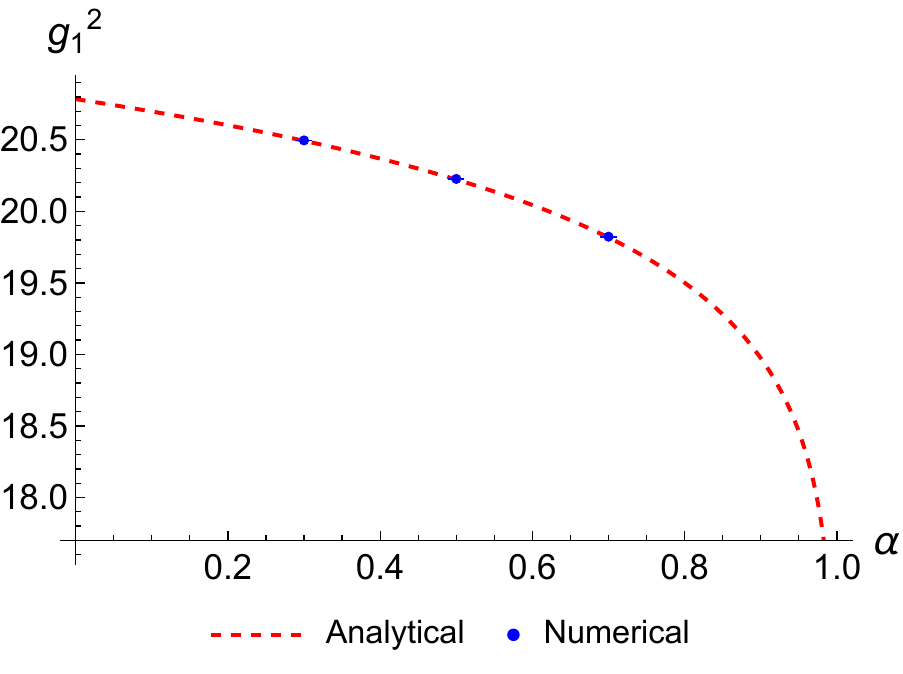}
	\caption[Numerical data vs. analytical, 2D inelastic dual]{Comparison between the analytical expression (\ref{eqn:analyticinelastic}) and the data for $\beta^{(\textrm{s})}$ with some tiny error bars.}
	\label{fig:inelasticdualvsana}
\end{figure}

\begin{figure}
	\centering
		\includegraphics[width=0.8\linewidth]{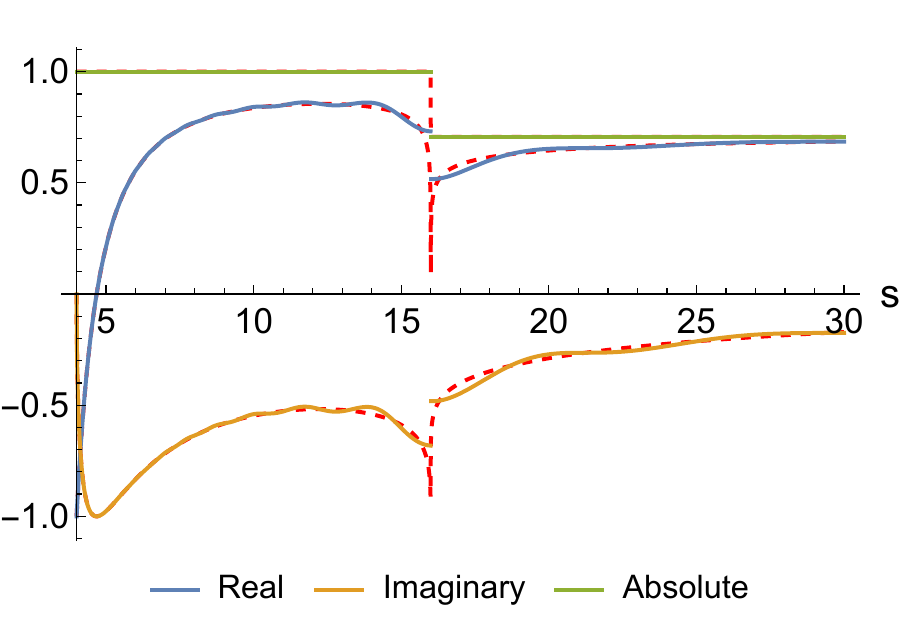}
	\caption[$1/N_{max}$ vs. $D$, 2D inelastic]{S-matrix components for the inelastic profile $\beta^{(\textrm{s})}$ for the dual approach vs. the analytical components, red dashed curves. $m_1^2 = 3$, $N_{max} = 40$, $\alpha = 0.5$ and $s^*=16$ were used.}
	\label{fig:compsmatdualinelastic}
\end{figure}

{\bf Dual in 2d.} 
In terms of optimization language, the primal problem is
\begin{equation}
	\underset{{\{T(s),\,g_1^2\}}}{\text{maximize}}\quad g_1^2
\end{equation}
subject to the constraints 
\begin{equation}
	\mathcal{A}(s) \equiv T(s)+\frac{g_1^2}{s-m_1^2}-\int_{4m^2}^{\infty}\frac{dz}{\pi}\frac{\text{Im} \,T(z)}{s-z}-(s \leftrightarrow t) = 0
\end{equation}
and the new unitarity condition, which is given by
\begin{align}
	\mathcal{U}(s) &\equiv 2\sqrt{s(s-4)}\left(1-\beta(s)\right) +2\,\text{Im}\,T(s) \nonumber \\
	&-\frac{1}{2\sqrt{s(s-4)}}\,|T(s)|^2 \geq 0\,.
\end{align}
To achieve a dual formulation, we closely follow \cite{Guerrieri:2020kcs} keeping track of the $\beta(s)$ dependence and build the Lagrangian density $\mathcal{L}(T,\omega,\lambda)$,
\begin{equation}
	\mathcal{L}(T,\omega,\lambda) = g_1^2+\int_{4m^2}^{\infty}ds\,\omega(s)\mathcal{A}(s)+\lambda(s)\mathcal{U}(s)\,,
\end{equation}
where $\lambda$ and $\omega$ are Lagrange multipliers and $\lambda \geq 0$. Using the weak duality principle and introducing the function $W(s)$ satisfying $\textrm{Im} W(s)= \omega(s)$, which does not depend on the form of the unitarity condition, we arrive at the expression
\begin{equation}
	\mathcal{L}(T,\omega,\lambda) = \int_{4m^2}^{\infty}ds\,\text{Im}\left[W(s)T(s)\right]+\lambda(s)\mathcal{U}(s)
\end{equation}
provided that $\text{Re}\,W(m_1^2)=-1/\pi$. After solving the equations of motion for $\lambda$ and $T(s)$,
we conclude that the correct dual optimization problem is
\begin{equation}
	\underset{{\{W\}}}{\text{min}}\,\, D(W) = \int_{4m^2}^{\infty}\frac{ds}{\rho_{11}^2(s)}\,\left[|W(s)|\beta(s)+\text{Re}\,W(s)\right]\,,
\end{equation}
where $\rho_{11}^{-2}=2\sqrt{s(s-4)}$
subject to the condition that 
\begin{equation}
	W(m_1^2)=-\frac{1}{\pi}\,,
\end{equation}
where $\beta(s)$ are the profiles used above.
Lastly, it is necessary to propose an ansatz to carry out the numerical extremization. The naive $\rho(s)$ expansion can still be used
\begin{equation}
	W(s) = \frac{1}{s(4m^2-s)}\sum_{n=1}^{N_{max}}a_n(\rho(s)^n-\rho(t)^n)\,,
\end{equation}
where $N_{max}$ is again the numerical cutoff. The fact that the same Ansatz was used even though the S-matrix encodes additional thresholds is remarkable when compared to the primal formalism. This follows from the fact that the absolute value of the S-matrix is fixed in the dual procedure because of the strong duality principle. Namely
\begin{equation}
	T(s) = \frac{i}{\rho_{11}^2(s)}\left(1+\frac{W^*(s)}{|W(s)|}\sqrt{\beta(s)}\right)\,,
\end{equation}
which has the correct absolute value regardless of $W$.
Revisiting the cubic coupling maximization yields figure \ref{fig:inelasticdualvsana}. Here we performed a simple extrapolation in $N_{\textrm{max}}$ which agrees with the analytic result to one part in $10^{-4}$ and estimated some (very small) error bars. 
A look at S-Matrix components, figure \ref{fig:compsmatdualinelastic}, confirms the expectations. The absolute value exactly matches the profile $\beta(s)$, and the real and imaginary parts nicely follow the expected singular behavior.

\begin{figure}
	\centering
	\includegraphics[width=0.9\linewidth]{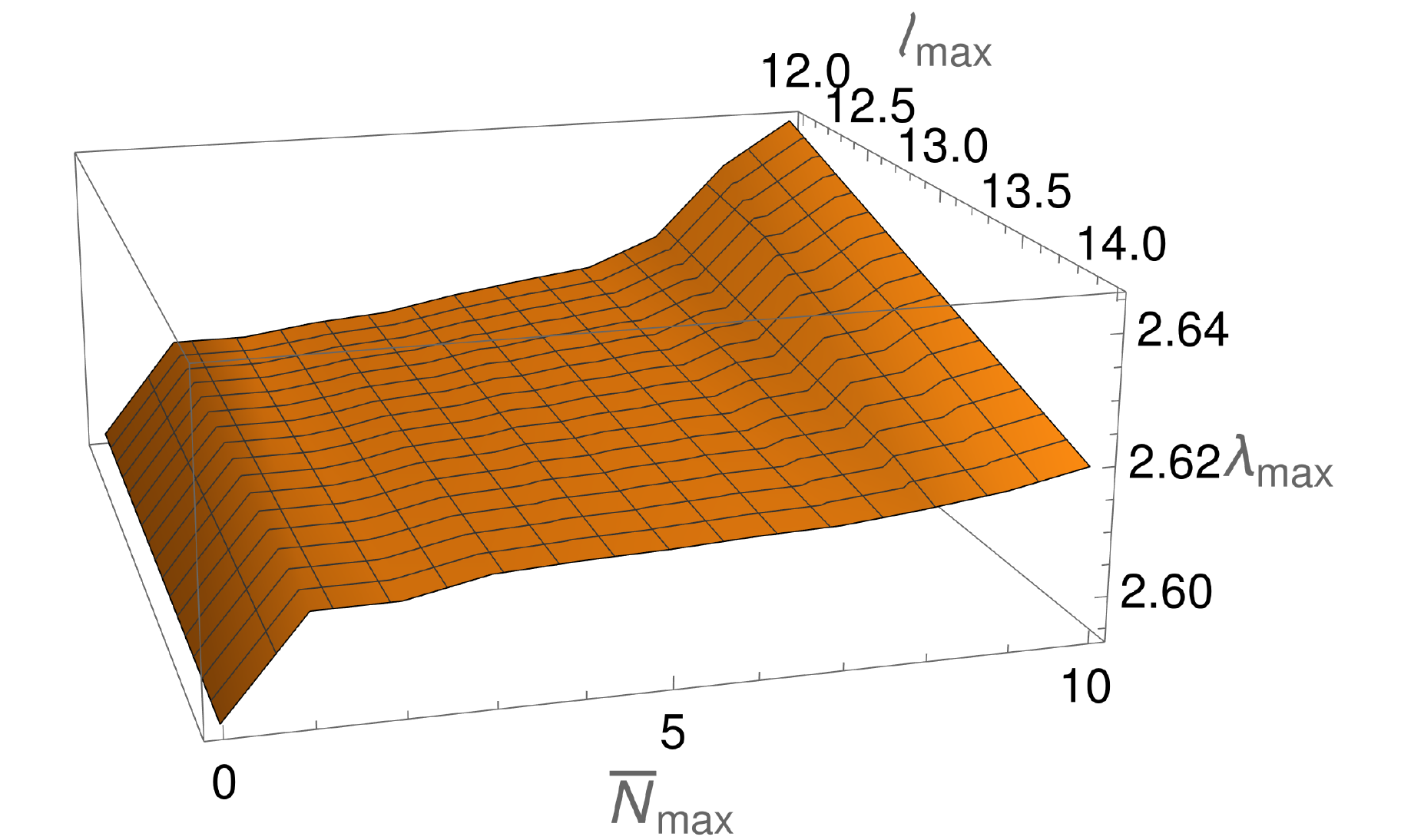}
	\caption[Quartic coupling maximization, 4D inelastic with $\rho_{16}$ and $\alpha = 0.2$]{Result of maximizing the quartic coupling subject to (\ref{mat:flconstraint}), for profile $\beta_0^{(\textrm{s})}(s)$ with $\alpha = 0.8$ and $s^*=16$. We find a plateau $\lambda \approx 2.61$ around $\bar{N}_{\textrm{max}} \approx 5$.}
	\label{fig:4d3dalpha02N14}
\end{figure}

\begin{figure}
	\centering
		\includegraphics[width=0.7\linewidth]{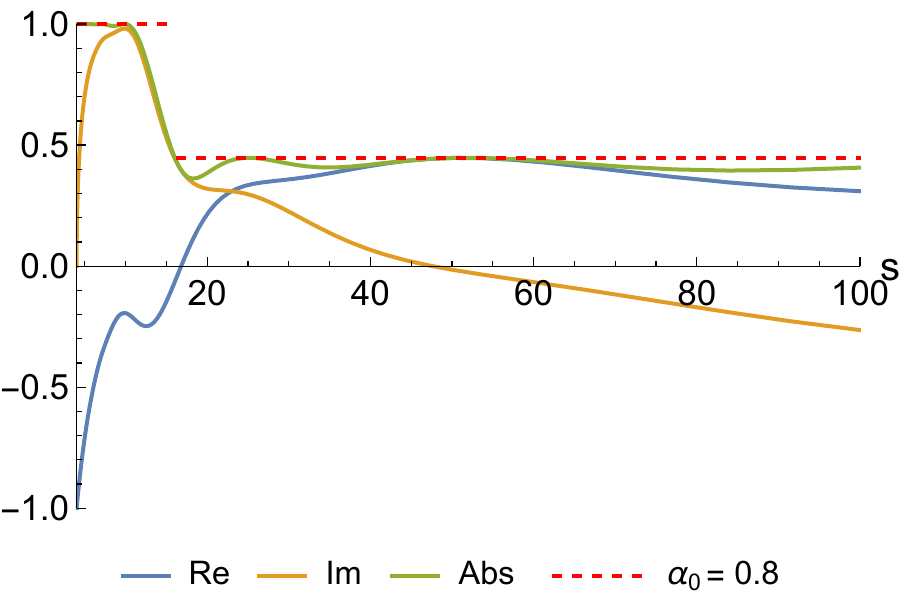}
		\centering
		\includegraphics[width=0.7\linewidth]{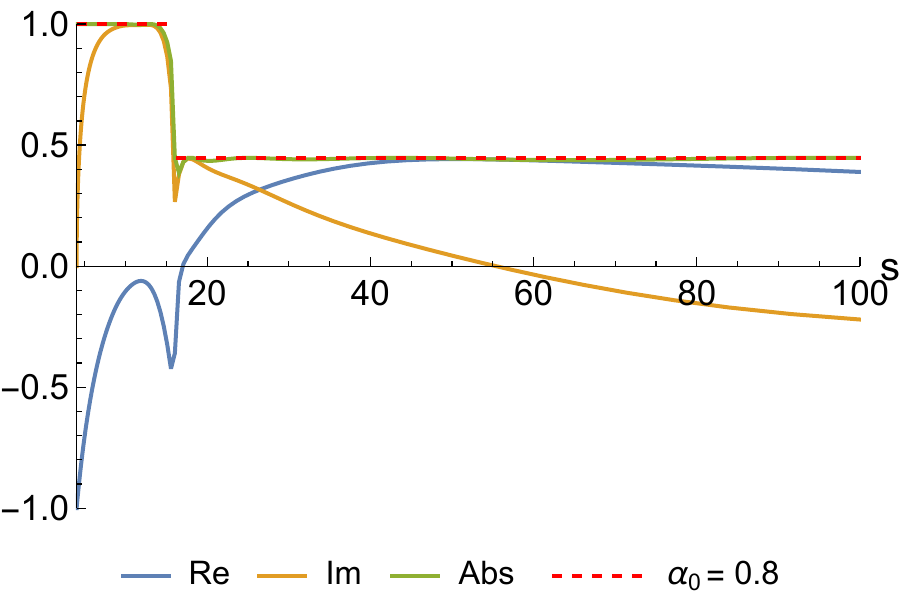}
	\caption[Quartic coupling maximization, 4D inelastic S-matrix components, $\alpha = 0.8$]{Spin zero partial amplitude maximizing the quartic coupling subject to \ref{mat:flconstraint}, for profile $\beta_0^{(\textrm{s})}(s)$, $N_{\textrm{max}} = 14$, $\ell_{\textrm{max}} = 14$ and $\alpha = 0.8$ with $\bar{N}_{\textrm{max}} = 0$ (top) and $\bar{N}_{\textrm{max}} = 9$ (bottom).}
	\label{fig:4drho1602comp}
\end{figure}

{\bf Primal in 4d.}
The natural way to generalize the inelastic constraints discussed in 2 dimensions above, is to restrict the absolute values of the partial amplitudes $S_\ell= 1 + i \sqrt{(s-4)/s}\, f_\ell$, where $f_\ell$ are the partial waves normalized as in \cite{Correia:2020xtr}. We then impose
\begin{equation}
\label{eq:inelasticpartialamplitudes}
	|S_\ell(s)|^2\leq \beta_\ell(s), \quad s\geq 4.
\end{equation}
A natural starting point would be choosing a profile compatible with the Roy-Martin production rate (\ref{eq:RoyMartin}). This is of course only asymptotic, but imposing it from moderate energies, we found the effect to be negligible. This is to be expected since this lower bound is exponentially small.
Instead, we will study step profiles, identical to the ones used in the previous sections. In practice, we will use non-trivial inelasticity profiles only for the low spin partial-waves and require standard unitarity for the remaining ones. To use (\ref{eq:inelasticpartialamplitudes}) in terms of semi-definite programming we rewrite it as a matrix inequality
\begin{equation}\label{mat:flconstraint}
\small
\begin{pmatrix}
1-\frac{\text{Im} \, a_\ell}{1+\sqrt{\beta_\ell(s)}} & \text{Re}\,a_\ell \\
\text{Re}\,a_\ell & \beta_\ell(s)-1 +\left(1+\sqrt{\beta_\ell(s)}\right)\,\text{Im}\,a_\ell\\
\end{pmatrix}
\succeq 0
\end{equation}
where $a_\ell=\sqrt{(s-4)/s}\,f_\ell$. In 4d, we will stick to the simplest maximization problem, which can be studied even for a model without bound states: Maximization of the quartic coupling $\lambda\equiv \Lambda_{00}/32\pi$. Motivated by the 2d explorations, we start from an ansatz
\begin{align}\label{eqn:4dansatzmodified}
T(s,t,u) =  \,\,&\kappa\left(\frac{1}{\rho_s-1}+\text{cross.}\right)+\sum_{a,b,c=0}^{N_{\textrm{max}}}\,\alpha_{abc}\,\rho_s^a\rho_t^b\rho_u^c \nonumber\\
&+\sum_{a,b,c=0}^{\bar{N}_{\textrm{max}}}\,\beta_{abc}\,\bar{\rho}_{s}^a\bar{\rho}_{t}^b\bar{\rho}_{u}^c
\end{align}
where we use the $\bar{\rho}$ variables discussed above, and add a pole at threshold, parametrized by $\kappa$, which is known to capture the physics of the extremal S-Matrix, in the standard case \cite{Paulos:2017fhb}. First, we consider a step profile for the spin zero partial wave $\beta_0^{(\textrm{s})}(s)$, with $\alpha=0.5$ and $s^*=16$. Even when we don't use the new terms in the ansatz, i.e. when $\bar{N}_{\textrm{max}}=0$, we find a substantial difference compared to the "elastic" case, as can be seen in figure \ref{fig:fl0}:
\begin{figure}
		\centering
		\includegraphics[width=0.75\linewidth]{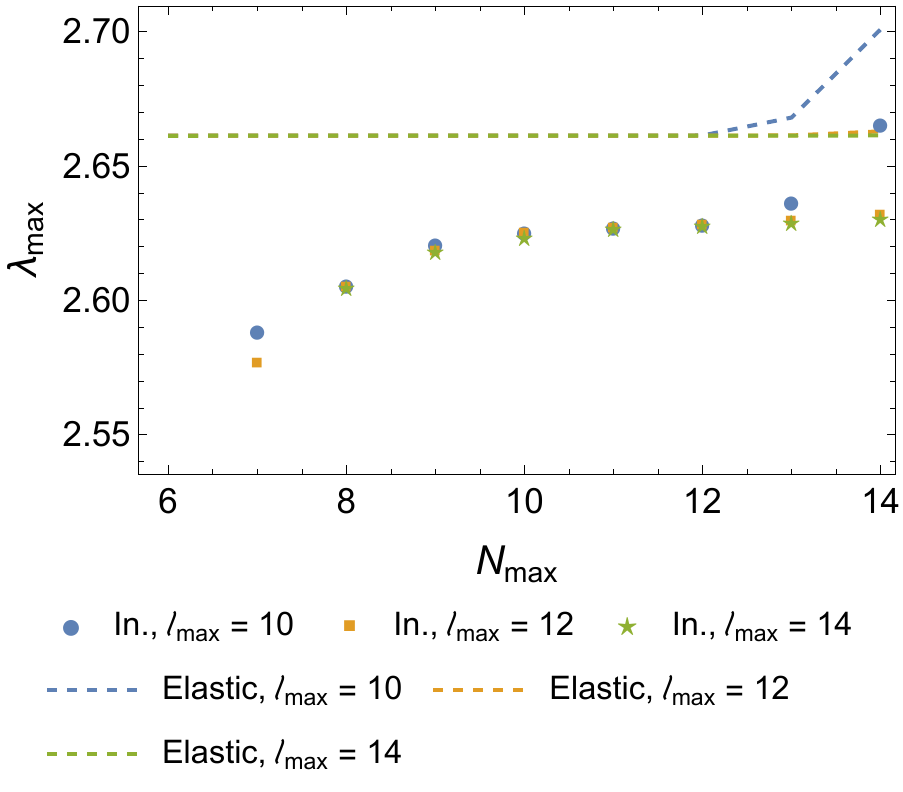}
	\caption[Quartic coupling maximization, 4D inelastic with restricted partial-waves]{Result of maximizing the quartic coupling subject to (\ref{mat:flconstraint}), for profile $\beta_0^{(\textrm{s})}(s)$ with $\alpha = 0.5$ and $s^*=16$. The dashed lines correspond to the solution with unitarity only.}
	\label{fig:fl0}
\end{figure}
It seems that $\lambda$ roughly achieves a plateau, which is clearly below the original bounds, where no inelasticity was imposed. However, extending the ansatz shows that this difference is somewhat overestimated. Varying the parameter $\bar{N}_{\textrm{max}}$, with $N_{\text{max}}=14$, one produces figure \ref{fig:4d3dalpha02N14}, where we considered a somewhat sharper discontinuity with $\alpha=0.8$.
 Clearly, the quartic coupling now plateaus at a slightly higher value, $\lambda_{\textrm{max}} \approx 2.61$ , which is quickly achieved by adding merely a few $\bar{\rho}$ terms. We can now also look at the partial amplitudes, and observe how the particle production is realized, as shown in figure \ref{fig:4drho1602comp}. Plotting the spin zero amplitude $S_0$, we can observe the important effect of the more general ansatz. Not only is the transition smoother and less oscillatory but the constraints are more sharply saturated, and in particular elastic unitarity is satisfied remarkably well.

\begin{figure}
	\centering
	\includegraphics[width=0.9\linewidth]{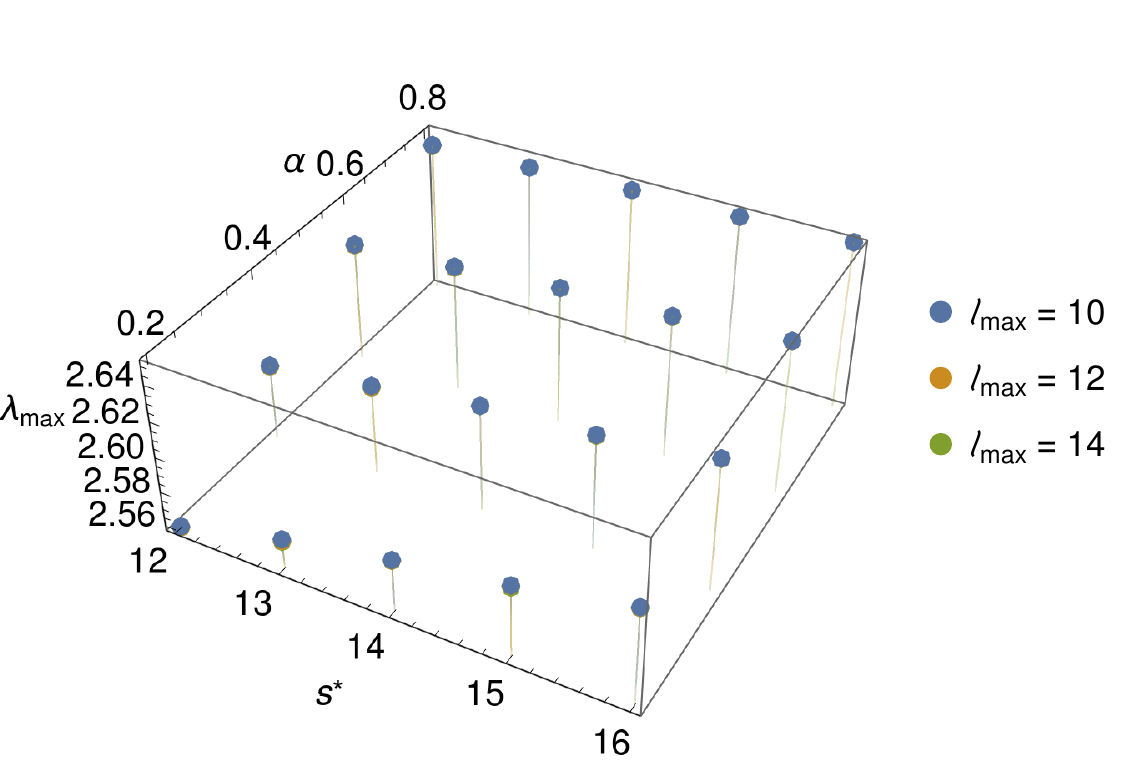}
	\caption{Maximum quartic coupling $\lambda_{\textrm{max}}$ as a function of $\alpha$ and $s^*$, for a profile $\beta_0^{(\textrm{s})}$. The results have converged since changing $\ell_{\textrm{max}}$ from 10 to 14 leaves the bounds unchanged. Here, $N_{\textrm{max}}=10$ and $\bar{N}_{\textrm{max}}=6$.}
	\label{fig:alfasstar}
\end{figure}

Having established the reliability of the new ansatz (\ref{eqn:4dansatzmodified}), we now give a more interesting application. For the profile $\beta_0^{(\textrm{s})}$, we study how the maximum quartic coupling varies as we tune both the strength of the discontinuity $\alpha$ and the start of the inelastic threshold $s^*$. The results are presented in figure \ref{fig:alfasstar} \cite{note3}. As expected, both making the discontinuity softer (decreasing $\alpha$), and pushing the inelastic threshold further way (increasing $s^*$) make the bounds approach from below the value obtained imposing simply unitarity, $\lambda_{\textrm{max}} \approx 2.66$. We also note that for this profile it is possible to accurately replicate very sharp discontinuities, as we find stable results, even for $\alpha$ as small as $0.01$.

As a final application of the methods we developed, we also consider the effect of inelasticity on the spin 2 partial wave $f_2$. Maximizing the quartic coupling $\lambda$ with profile $\beta_2^{(\textrm{s})}$ with parameters $\alpha=0.8$ and $s^*=16$ leads to a bigger decrease in the coupling than the corresponding spin 0 problem. However, convergence is worse and a rough plateau is only achieved for larger values of $\bar{N}_{\textrm{max}}$. Concretely, the maximum value of the coupling seems to drop considerably to about $\lambda_{\textrm{max}}\approx 2.5$. This is perhaps unsurprising as the contribution of a spin $\ell$ partial wave to the amplitude is weighted, in our normalization, by a factor of $2\ell+1$ (the same factor appears below in the contribution to the total cross section (\ref{eq:totalcross})), effectively enhancing the spin 2 contribution by a factor of 5. An interesting question for the future is to understand what happens in the physical cases where the inelasticity decays asymptotically with spin, as in the black-hole production case, or with Dragt behavior \cite{Dragt:1967}. Such an understanding would require a reliable way to estimate the finite spin effects.

\begin{figure}
	\centering
	\includegraphics[width=0.9\linewidth]{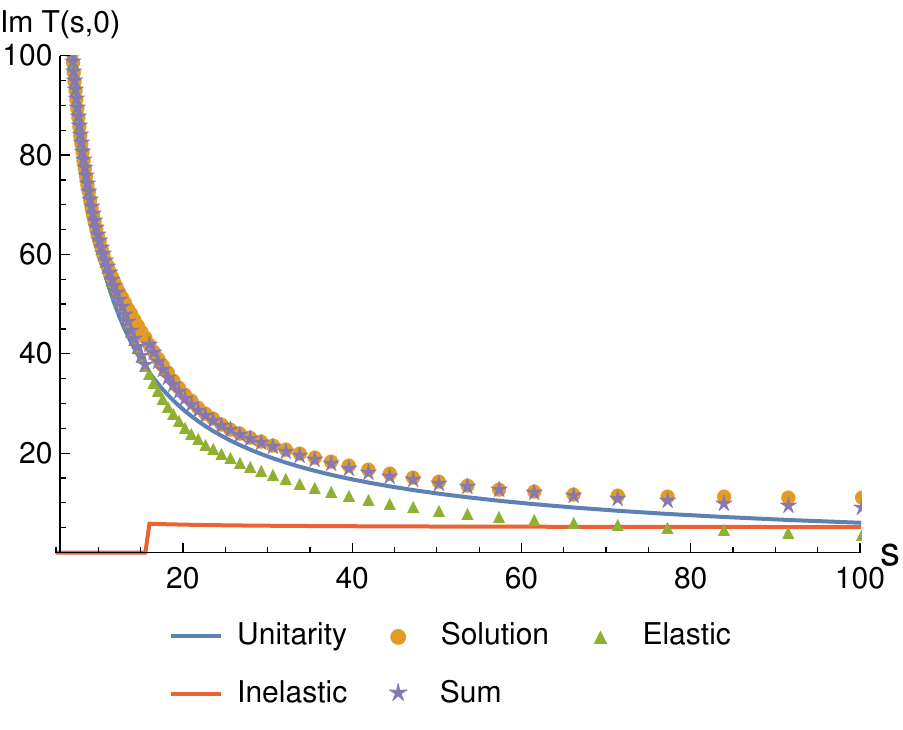}
	\caption[Comparison of various plots, $f_{\ell}$]{Comparison between the solution where only unitarity is imposed, \textit{Unitarity}; the numerical solution with inelasticity imposed in $S_0$, \textit{Solution}; the elastic part of the amplitude obtained from the optical theorem, \textit{Elastic}; the imposed inelastic profile, \textit{Inelastic}; and the sum of elastic and inelastic components \textit{Sum}.}
	\label{fig:flcompare0.2}
\end{figure}

{\bf Total Cross-section.}
Having obtained stronger bounds on the couplings and captured the extra threshold, we proceed to analyze the total cross-section $\sigma_t$, or equivalently the forward limit $T(s,0)$ of the resulting amplitude. The relation between the two is simply
\begin{equation}
\label{eq:totalcross}
\text{Im}\,T(s,0) = \sqrt{s(s-4)}\,\sigma_{t}=\sum_{\ell}16\pi(2\ell+1)\text{Im}f_\ell\,,
\end{equation}
and of course, we can decompose the cross-section into its elastic and inelastic parts. Notably, we can use the optical theorem to obtain the elastic part of the amplitude in the forward limit in terms of an integral of its square
\begin{equation}
\text{Im}\,T(s,0) \geq \frac{1}{64\pi}\sqrt{\frac{s-4}{s}}\int_{-1}^{1}dz\,\left|T(s,t(z))\right|^2\,,
\end{equation}
with $z$ the cosine of the scattering angle.
The inequality is saturated in the elastic unitarity region, and after that there are inelastic terms further contributing. This can also be used as a check to the validity of the bounds obtained when imposing inelasticity.
We plot the forward amplitude corresponding to maximizing the quartic coupling with $\alpha=0.2$ and $s^*=16$ in figure \ref{fig:flcompare0.2}. We also computed the integral in the optical theorem numerically, and determined the inelastic contribution to the total cross section through the sum over partial waves in (\ref{eq:totalcross}). We find good agreement between the sum of the elastic and inelastic pieces and the resulting cross-section obtained directly through the maximization process, as shown in figure \ref{fig:flcompare0.2}. 

\begin{figure}
	\centering
	\includegraphics[width=0.9\linewidth]{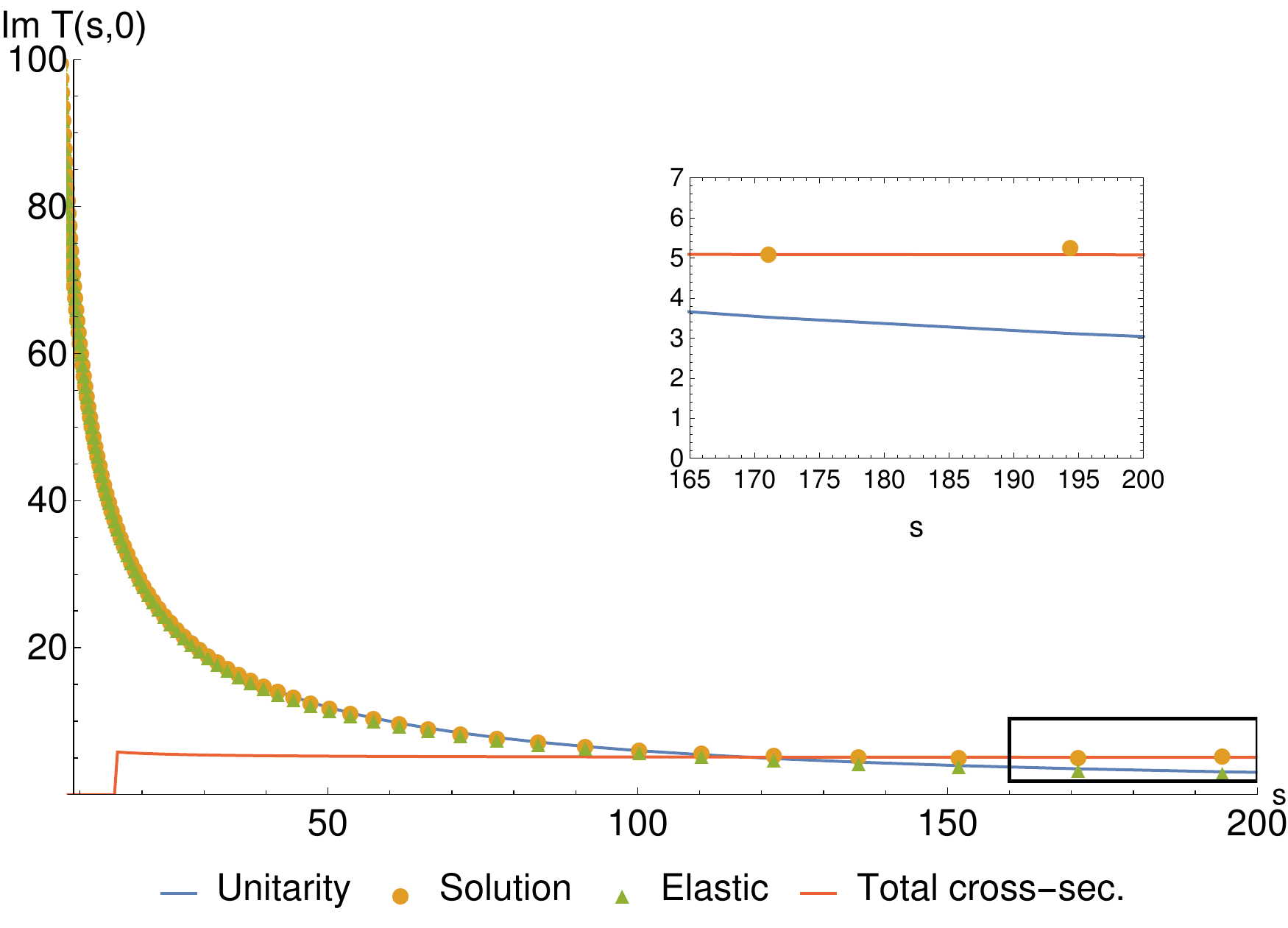}
	\caption[Comparison of various plots, $"\sigma_{total}"$]{Comparison between the solution with only unitarity imposed, \textit{Unitarity}; the solution with $\sigma_{t}$ that follows from $\beta_0^{(\textrm{s})}(s)$ with $\alpha = 0.2$, \textit{Solution}; the elastic part of the amplitude from the optical theorem, \textit{Elastic}; and the imposed cross-section \textit{total cross-sec.}}
	\label{fig:pseudosigmacompare0.2}
\end{figure}

We find that the forward amplitude always exceeds its elastic counterpart denoted by the solid blue line, which is obtained by maximizing the quartic coupling imposing only unitarity. Indeed, this also suggests a natural way to parametrize the inelasticity: inputing a lower bound on the total cross-section, which is a simple generalization of imposing positivity in the forward limit. This constraint mixes information about all the partial waves, so it should lead to weaker bounds than imposing inelasticity for each partial wave as in (\ref{mat:flconstraint}). In practice, to fully control the inelasticity, one needs to include the quadratic elastic piece which leads to a constraint that is hard to use. On the other hand, one can impose
\begin{equation}\label{mat:RMconstraint}
\begin{pmatrix}
\text{Im}\,T(s,0) - \sqrt{s(s-4)}\,\sigma_{t} & 0 \\
0 & 1\\
\end{pmatrix}
\succeq 0
\end{equation}
where $\sigma$ is just treated as a fixed function of $s$. In practice, this just leads to saturation of this profile once the elastic cross-section exceeds this value, as shown in figure \ref{fig:pseudosigmacompare0.2}.
We also point out that due to the nature of the $\rho$ expansion, the forward limit of the amplitude always decays at large $s$. It would be nice to study extensions which allow for Froissart-like behavior.

{\bf Discussion.}
In this paper, we explored how inequalities stronger than unitarity, i.e. imposing an inelasticity profile, yields stronger bounds for the low-energy observables one studies with the S-matrix bootstrap. To correctly capture the new thresholds, we introduced the $\bar{\rho}$ variables, which drastically improve convergence and allowed for the satisfaction of elastic unitarity. As expected, we qualitatively found that the maximum coupling decreases as we increase the amount of inelasticity, which can be achieved by making the discontinuity sharper or by lowering the energy of the first inelastic threshold, as quantified in figure \ref{fig:alfasstar}. We also found that the same amount of inelasticity imposed on a higher spin partial wave further decreases the maximum coupling, which is kinematically expected.

More quantitatively, we found that the minimal particle production rate predicted by Roy and Martin \cite{Martin:2017ndt} has a negligible effect on the quartic coupling bounds, but that step-like profiles which emulate the basic features of particle production \cite{Zamolodchikov:2011wd} lead to modifications on the bound on quartic couplings of the order of a few percent. Furthermore, the introduction of the $\bar{\rho}$ variables alters these values substantially, and their use seems crucial for precision studies (say once actual experimentally/theoretically motivated profiles are added to the problem). A flagship application to consider is the study of supersymmetric gravity amplitudes in higher spacetime dimensions \cite{Guerrieri:2021ivu,Guerrieri:2022sod}. In 10 dimensions, these authors found that the space of allowed couplings is almost completely spanned by the predictions of string theory as one varies the string coupling, but a difference of a few percent exists between the numerical minimum and the string theory prediction. Adding an inelasticity profile (in that case derived from the Bekenstein-Hawking formula) combined with the methods of this paper might slightly modify the numerical minimum, as we found in our toy models, perhaps conclusively showing saturation of the bound by string theory.

Finally, our analysis of the cross-section suggests that it could be possible to obtain inelastic S-matrices by maximizing the total cross-section. Of course, to maximize the total cross-section at a fixed value of the energy is not expected to be a physically well posed problem. The existence of resonances allows for sharp peaks in the cross-section, which are obviously physical and should not be excluded. A naive use of the maximization methods, leads to results that do not seem to converge as one increases $N_{\textrm{max}}$, confirming the above expectation. Curiously, for fixed $N_{\textrm{max}}$, and varying $s$, the values seem to roughly follow a Froissart-like trend. It would be interesting to understand how this would change for ans\"atze that allow for amplitudes that grow at infinity.
On the other hand, suitable integrals of the cross section over energy should eliminate the problem mentioned above, and might prove to be an interesting observable.
\linebreak

	We thank C. Bercini, M. Correia, V. Goncalves, A. Guerrieri, A. Hebbar, A. Homrich, J. Penedones and P. Vieira for illuminating discussions and useful comments and M. Correia and A. Hebbar for useful comments on the draft. This  research  received  funding  from  the  Simons Foundation grant 488637 (Simons collaboration on the non-perturbative bootstrap). AA received funding from the German Research Foundation DFG under Germany’s Excellence Strategy – EXC 2121 Quantum Universe – 390833306.  Centro  de  F\'isica  do  Porto  is  partially  funded  by  Funda\c{c}\~ao  para  a  Ci\^encia  e  Tecnologia (FCT) under the grant UID04650-FCUP.
  
\begin{figure}[ht]
	\centering
	\includegraphics[width=0.75\linewidth]{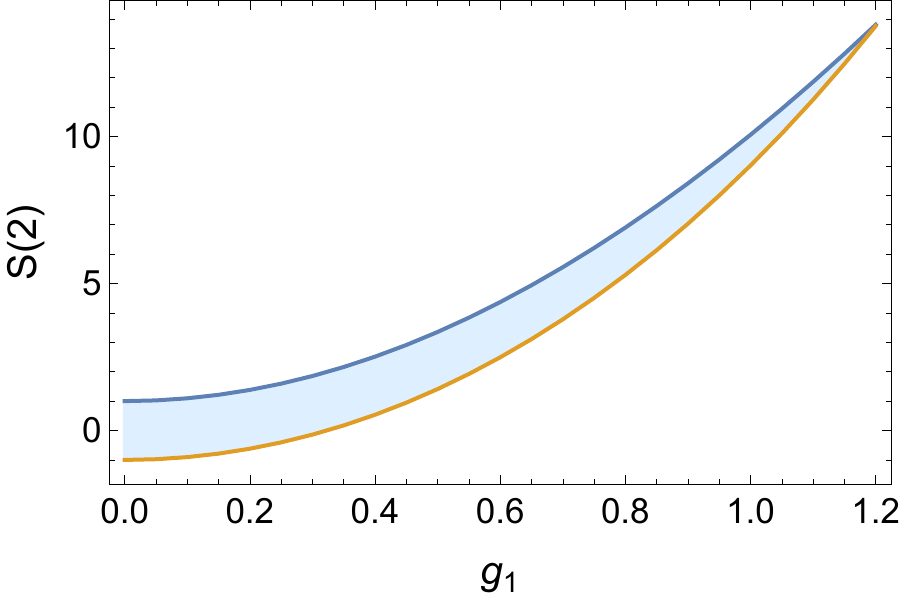}
	\caption[$S(2)$ vs. $g_1^2$, 2D]{Plot of $g_1$ vs. $S(2)$ for $m_1 = \sqrt{3}$. The blue curve represents the maximum value of $S(2)$ while the orange represents the minimum.}
	\label{fig:S2vsg}
\end{figure}

{\bf Appendix A.}
In the main text, we used inelasticity as a way to probe the interior of the bootstrap bounds. Here, we study a simpler problem which also allows us to visit non-extremal theories with respect to maximizing the cubic coupling. We fix the cubic coupling to be $0<g_1<g_{\textrm{max}}$, and then maximize and minimize the quartic coupling $S(2)$. The result is figure \ref{fig:S2vsg}. At the tip of maximum coupling, the solution is unique and is saturated by sine-Gordon. In the other extreme $g_1=0$, the bound state decouples, and the maximum and minimum theories are just a free boson and fermion respectively. In between, there are non-trivial solutions, generated by products of CDD factors. One of the solutions can be seen as adding the $T\bar{T}$ deformation in the vicinity of the sine-Gordon point.

\begin{thebibliography}{99}
\bibitem{note1} Strictly speaking, the $\bar{\rho}$'s can also be expressed in terms of the ordinary $\rho$'s but this requires infinitely many terms, and therefore including them substantially improves the convergence.
\bibitem{note3} For this plot we used only $\bar{\rho}$ variables with threshold at $s^*=16$, to avoid recomputing expensive partial wave integrals. Minor changes are to be expected if one adjusts the $\bar{\rho}$ variables for the respective value of $s^*$.
\end{thebibliography}
\end{document}